\documentclass[prb,showpacs,twocolumn]{revtex4}
\usepackage{dcolumn}
\usepackage{bm}
\usepackage{amssymb}
\usepackage{amsmath}
\usepackage{amsfonts}
\usepackage[]{graphicx}
\begin{document}
\bibliographystyle{prsty}
\title{Quantum plasmonics: second-order coherence of surface plasmons launched by quantum emitters into a metallic film}
\author{Oriane Mollet,$^{1}$ Serge Huant,$^{1}$ G\'{e}raldine Dantelle,$^{2}$ Thierry Gacoin,$^{2}$ and Aur\'{e}lien Drezet$^{1}$}
\affiliation{$^{1}$Institut N\'eel, CNRS, and Universit\'e Joseph Fourier, BP 166, 38042 Grenoble, France}
\affiliation{$^{2}$ Laboratoire de Physique de la Mati\`ere Condens\'ee, Ecole Polytechnique, UMR CNRS 7643, 91128 Palaiseau, France}

\date{\today}

\begin{abstract}
We address the issue of the second-order coherence of single surface plasmons launched by a quantum source of light into extended gold films. The quantum source of light is made of a scanning fluorescent nanodiamond hosting five nitrogen-vacancy (NV) color centers. By using a specially designed microscopy that combines near-field optics with far-field leakage-radiation microscopy in the Fourier space and adapted spatial filtering, we find that the quantum statistics of the initial source of light is preserved after conversion to surface plasmons and propagation along the polycrystalline gold film. \\

 \end{abstract}


\maketitle
\section{Introduction}
Quantum plasmonics, the study of surface plasmons (SPs) at the quantum level, is a promising framework for building new technological platforms at the nanoscale. As hybrid photon-electron states naturally confined at a metal-dielectric boundary,~\cite{Barnes} SPs share fundamental properties of both their electronic and photonic faces~\cite{Novotny} that are specifically modified in the quantum regime. From the electronic side first, confinement and hybridization can generate new cooperative electronic states between two or more metal particles,~\cite{Maierbook} whereas in the contact plasmonic regime, the quantum nature of the electron cloud can come into play.~\cite{Maier} This will impact on various applications such as surface enhanced Raman scattering or non-linear optics that rely on a tight control over dimensions and distances.~\cite{Maier} Second, the quantum optical face of the SPs, which is the main focus of the present paper, draws a promising avenue for quantum information processing or cryptography in 2D, despite intrinsic limitations due to the finite propagation length. The control over the particle-like part of the SPs is a necessary prerequisite for quantum transfer of information, which could lead to secure information routing along metal nanowires as well as qubit entanglement mediated by SPs.~\cite{Chang,Tudela,Woerdman,Gisin}\\
\begin{figure}[hbtp!]
\begin{center}
\begin{tabular}{c}
\includegraphics[width=7.5cm]{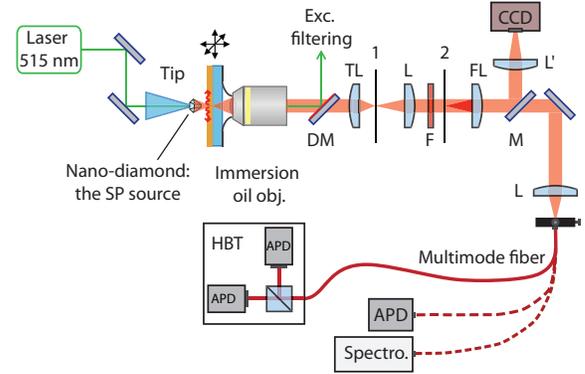}
\end{tabular}
\end{center}
\caption{Sketch of the experimental setup. The X100 oil immersion objective is used with a matching oil of optical index $n_o=1.46$. The effective numerical aperture is $NA_{\textrm{eff}}=1.35$. TL is the tube lens of the inverted microscope. The L and FL (like Fourier lens) doublet lenses have the same focal length 7.5 cm and the L' lens has a focal length of 15 cm. Planes 1 and 2 stand for the direct and Fourier space imaging modes with the CCD (charge-coupled device) camera, respectively. M is a switchable mirror used for selecting either the upper detection path, that offers a direct imaging on a CCD camera, or the lower detection path. The latter comprises a NSOM imaging branch on an avalanche photodiode (APD), a Hanbury-Brown \& Twiss (HBT) interferometer for photon correlation measurements, and a spectrometer for spectral analysis. A plug-and-play selection between the three branches is ensured by a multimode fiber in order to keep the same optimal conditions during the whole study. A dichroic mirror DM and a high-pass filter F cut the remaining excitation and spurious light below $\lambda=578$ nm, which mostly comes from the spurious tip and substrate emission. }
\end{figure}
\indent One key issue however is the preservation of quantum coherence throughout the interaction of light with the metal. Ohmic losses and metal roughness are known to degrade the first-order coherence (fringe visibility) of propagating SPs.~\cite{Jo} To what extent this will affect the second-order, i.e., quantum coherence, has so far only been investigated in a few specific situations. For instance, it has been shown that the photon correlation in an entangled pair generated by parametric down-conversion can be preserved after interaction with a plasmonic hole array or metallic stripe waveguides.~\cite{Woerdman,Gisin,Dimartino} Other studies focused on the generation of single SPs using single-photon emitters coupled to a nanowire~\cite{Lukin} or an antenna.~\cite{Benson1} It was found experimentally that the antibunching dip in the second-order time correlation of the single photon intensity is preserved even after optical coupling through the plasmonic nanowire or particle,~\cite{Fedutik,Benson2,Kolesov,Huck} although the depth of the antibunching dip is affected by the metal fluorescence.~\cite{Benson2} However, these studies focused on single-crystalline metal structures obtained by bottom-up colloidal synthesis.~\cite{Jo} For technological applications, it is essential to obtain a control over second-order coherence during the SP propagation along less demanding polycrystalline structures, manufactured using traditional top-down approaches such as e-beam lithography of evaporated metal films, where de-coherence can occur through enhanced Ohmic losses or scattering at the grain boundaries in the rough metal.~\cite{Simon}  The building block in this context would be a quantum emitter coupled to a plain metal thin film. Such a simple system should be able to support the propagation of SPs over large distances and preserve the second-order correlation of the quantum emitter. This would open the door to integrated quantum plasmonics, where the quantum source of light, patterned plasmonic waveguides and other necessary devices~\cite{drezet-nl2007,Natphoto} will all be integrated on a single chip.\\
\indent In the present paper we study such a building block by using a specially designed method that combines near-field scanning optical microscopy (NSOM)~\cite{Novotny,Brun,DREZET2002} with leakage radiation microscopy (LRM)~\cite{Hecht,Drezet1} in the Fourier space and adapted spatial filtering. As quantum emitters we consider nitrogen vacancy (NV) color centers hosted in a diamond nanocrystal.~\cite{Sonnefraud} Due to their remarkable photostability at room temperature and their large quantum efficiency, NV centers have already demonstrated their potential in quantum plasmonics~\cite{Benson1,Benson2,Kolesov,Cuche3} and consequently form ideal candidates
\begin{figure}[h]
\begin{center}
\begin{tabular}{c}
\includegraphics[width=7.5cm]{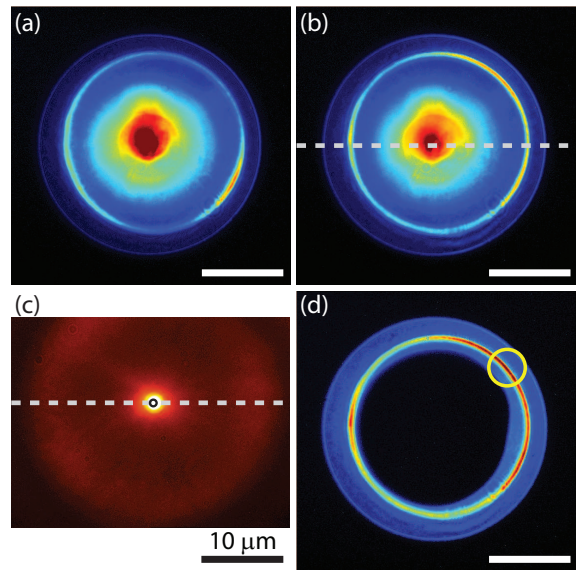}
\end{tabular}
\end{center}
\caption{(a)-(b) LRM images in the Fourier plane taken before and after the ND grafting by the NSOM tip, respectively. The white line in (b) is the crosscut direction analyzed in Figure 3(a). (c) LRM image in the direct plane taken after the ND grafting. The small black circle at the center of the figure corresponds to the multimode fiber detection area with a 50$\mu$m diameter multimode fiber after correction for the magnification of the setup (see text). The white line is along the crosscut direction analyzed in Figure 3(b). (d) Same as (b) with an optical mask placed in the Fourier plane. The yellow circle crossing the SP ring corresponds to the detection area with a 365 $\mu$m diameter multimode fiber. The scale bars in (a), (b) and (d) correspond to $NA=1$.}
\end{figure}
for the present study. We consider here a single fluorescent nanodiamond (ND) attached to the apex of a NSOM probe~\cite{Cuche,Cuche2} that serves the purpose of a scanning point-like quantum source of light.  Our motivation, is that the coherence properties of a single quantum source of light can be studied in a single run just by positioning this source in various environments with nanometer accuracy in all three dimensions (one capability of NSOM), i.e., in front of a gold film or in front of a silica substrate. This mimics a quantum emitter integrated in a final device. Moreover, the selected ND hosts a rather large number of NVs, in practice five, which allows to probe the coherence over many emitters at once. We show here that this many-NV ND still preserves its second-order coherence through the quantum plasmonic channel, thereby opening promising avenues for practical applications in integrated quantum plasmonics.\\
\section{Experimental details}
\indent Preparation of the ND sample was achieved following a procedure reported previously.~\cite{Dantelle2010,Rondin2010} To summarize, commercially available diamond nanocrystals are first irradiated using high-energy electrons, then annealed at 800 $^\circ$C to produce the fluorescent NV centers, and finally annealed in air at 550 $^\circ$C to remove surface graphitic compounds. Colloidal dispersion in water and further size selection allows obtaining a solution of mono-dispersed NDs. We consider here NDs with a typical size of 27 nm deposited on a 30 nm thin gold film evaporated on a fused silica substrate. For some measurements, part of the substrate is kept free from metal. As a preliminary test, the NV fluorescence is recorded using the scanning confocal mode of our microscope to locate isolated NDs on the gold film (see refs.~\cite{Sonnefraud,Cuche3,Cuche,Cuche2} for details). The principle of the plasmonics experiment is sketched on Figure~1, details will be commented in due place in the paper. In brief, we switch our microscope in the NSOM mode and use a dielectric NSOM tip.~\cite{Cuche3,Cuche,Cuche2} A laser light (wavelength $\lambda_{\textrm{exc}}=515$ nm) falling within the absorption band of the NVs is guided through the fiber to the tip apex where it excites the NV fluorescence. This will later be used to launch SPs in a gold film. The tip is then scanned over the region of interest, in the vicinity of a single ND elected for its brightness, and subsequently approached to the surface to graft the ND for further analysis and quantum plasmonics experiments. The grafting procedure has been detailed previously.~\cite{Cuche} Below, we comment in detail the different steps that demonstrate the conservation of the second-order coherence of SPs propagating along the gold film. \\
\section{Leakage radiation imaging of surface plasmons launched by single quantum emitters}
\indent Our demonstration makes extensive use of the LRM scheme.~\cite{Hecht,Drezet1} This far-field imaging method uses the ability of SPs propagating at the air-gold interface to leak into the substrate by tunnel effect through the (sufficiently thin) gold layer. Propagating SPs are characterized by an in-plane wavevector $K_{SP}(\lambda)=K'_{SP}(\lambda)+iK''_{SP}(\lambda)$, which is wavelength dependent and is a solution of an implicit dispersion relation.~\cite{Novotny,Drezet2} The real part $K'_{SP}(\lambda)$ can be written $2\pi n_{SP}(\lambda)/\lambda$, where we introduce the SP effective index $n_{SP}(\lambda)\gtrsim 1$. Therefore, phase matching at the different interfaces imposes the typical angle $\Theta_{SP}(\lambda)$ at which radiation leaks into the substrate and we obtain~\cite{Drezet2}
\begin{equation}
n\sin{(\Theta_{SP}(\lambda))}\simeq n_{SP}(\lambda)
\end{equation}
where $n\simeq 1.46$ is the optical index of fused silica. Since $\Theta_{SP}$ is larger than the critical angle $\Theta_{C}\simeq 43.2^{\circ}$ ($n\sin{(\Theta_{C})}=1$), LRM requires an immersion oil objective with $NA>1$. LRM can be used to study the SP propagation over long distances in the direct space. Additionally, it can be used for imaging the leaky waves in the Fourier or momentum space by recording the intensity in the back focal plane of the objective.~\cite{Hecht,Drezet1} The present optical setup allows the imaging of SPs either in the direct or Fourier space with a CCD camera by simply removing the Fourier lens (labeled FL in Figure 1).\\
\indent Figure 2(a) shows an LRM image in the Fourier plane obtained through the gold film prior to grafting the single ND. Here the NSOM tip is maintained at a $\simeq 20$ nm height above the ND. Due to adapted spectral filtering (Figure 1), the detected signal is essentially due to the NV and gold fluorescence (this is confirmed by the fluorescence spectra shown later on). The central spot in Figure 2(a) corresponds to the fluorescent light emitted in the `allowed' cone below $\Theta_{C}$, whereas the ring-like shaped distribution of intensity in the Fourier plane is typical for the SP launching by the NVs and corresponds to leaky spread at the angle $\Theta_{SP}$, i.e., in the `forbidden' region of the Fourier space. This observation confirms that the NV fluorescence emanating from a single ND deposited on a gold film is able to launch SPs.~\cite{Cuche3}\\
\begin{figure}[h]
\begin{center}
\begin{tabular}{c}
\includegraphics[width=8.2cm]{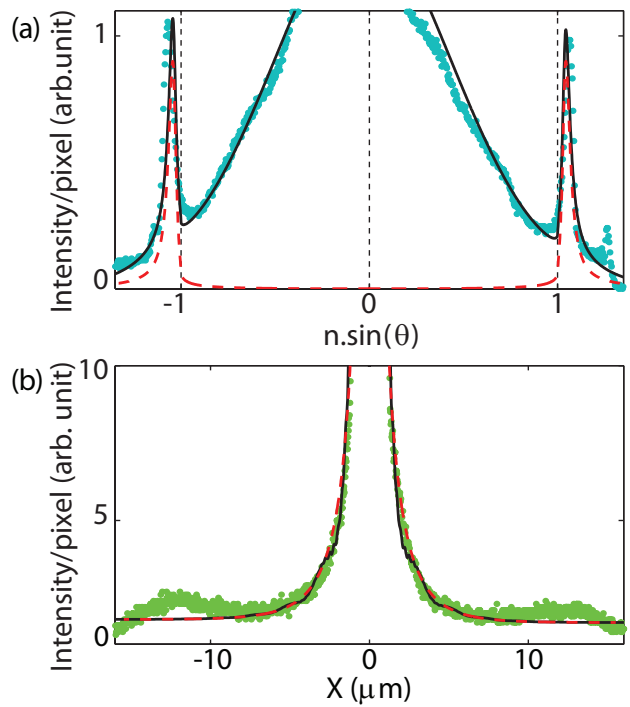}
\end{tabular}
\end{center}
\caption{(a) Intensity crosscut profile in the Fourier plane taken along the white dashed line diameter shown in Figure 2(b). The red curve shows the theoretical SP intensity profile for a transition dipole perpendicular to the metal film (see text). The total theoretical intensity including the nonplasmonic part of the signal is shown as the black curve. The small peak on the right hand side is the detection cutoff at $NA_{eff}\simeq 1.35$. (b) Intensity crosscut profile in the direct plane taken along the white dashed line diameter shown in Figure 2(c). The red curve shows the theoretical SP profile whereas the black curve includes the non-SP contribution as well. The small shoulders at large distances are attributed to the finite field of view of the microscope.}
\end{figure}
\indent Now, this very ND can be attached by the NSOM tip following a scheme that we described previously.~\cite{Cuche3,Cuche,Cuche2} Once the ND attached, plasmonics experiments can proceed further with the NV-based tip acting as source of light. The corresponding LRM images are shown in Figures 2(b-d). The Fourier-space image of Figure 2(b) depicts the same general features as prior to the ND grafting. A SP circle is clearly visible with some intensity anisotropy along the rim. This angular variation amounts to a relative ratio $(I_{max}-I_{min})/I_{max}\simeq 0.53$. It is likely due to the multi-dipolar nature of the point-like source that hosts several NVs (five, see below) and therefore two or three times more transition dipoles.~\cite{vincent} For completeness, we also show in Figure 2(c) an LRM image obtained in the direct space. The SP propagation is clearly visible as a diffused signal extending over several tens of microns. Note that the bright spot at the center of this image originates both from the plasmonic signal and from the intense non-plasmonic light transmitted directly through the thin metal film (allowed light).\\
\indent Before probing the correlation, i.e. particle-like properties of the SP signals shown in Figure 2(b,c), it is worth characterizing their wave-like properties through a quantitative analysis of the LRM images. For this purpose, we show in Figure 3(a) a crosscut along the SP diameter indicated by a dashed white line in Figure 2(b). From the SP peak position and using Eq.~1, we find the SP effective index $n_{SP}\simeq 1.06-1.07$ to be in close agreement with the value $n_{SP}(\lambda)\simeq 1.05$ extracted from the thin film dispersion in the spectral range of interest.~\cite{Drezet2} The width at half maximum of the SP peak in the Fourier space $W_{SP}(\lambda)$ gives the inverse of the SP propagation length, which for a monochromatic light reads~\cite{Drezet2} $L_{SP}=(2K''_{SP})^{-1}$. Here however, the NV emission is broad. Taking into account this broadening, the theoretical width can be estimated at $\langle W_{SP}\rangle/K'_{SP}\simeq 0.065 $, which is very close to the measured value $W_{SP}/K'_{SP}\simeq 0.07$. This implies propagation lengths varying between $L_{SP}\simeq 1.5$ and 3.5 $\mu$m in the NV emission band. \\
\indent Fig.~3(a) shows also the theoretical (normalized) SP peaks~\cite{Drezet2,Hohenau} obtained for a vertical dipole located at 20-30 nm distance from the gold film (red curve). Similar profiles are obtained for a horizontal dipole. It is worth pointing out that this good agreement between experimental and theory justifies our interpretation in terms of SPs propagating mainly at the metal-air interface: SPs propagating at the quartz-metal interface play no significant role and contribute only for a tiny fraction to the background signal observed at the center of the Fourier space image.~\cite{Drezet2}  \\
\indent The large background signal that dominates the allowed light zone below $\Theta_{C}$ is modeled by a Gaussian distribution. Clearly, the sum of both contributions (black curve) is in very good agreement with the experimental data. For completeness, we can also analyze the SP propagation in the direct space along a crosscut as shown in Figure 3(b). The experimental data are modeled with the total image intensity after propagation through the microscope (black curve). We add on the same graph the asymptotic SP intensity profile~\cite{Hecht,Brun,Drezet2} (red curve) which reads $I_{SP}(r)\propto e^{-r/L_{SP}}/r$, where $r$ is the distance between the in-plane observation point and the dipole position projection on the interface. By using the typical propagation length $\simeq 2.5$ $\mu$m obtained in Fourier space, a very good agreement is achieved, confirming therefore the SP contribution to the observed signal in both the direct and Fourier spaces. \\
\begin{figure}[h]
\begin{center}
\begin{tabular}{c}
\includegraphics[width=7.5cm]{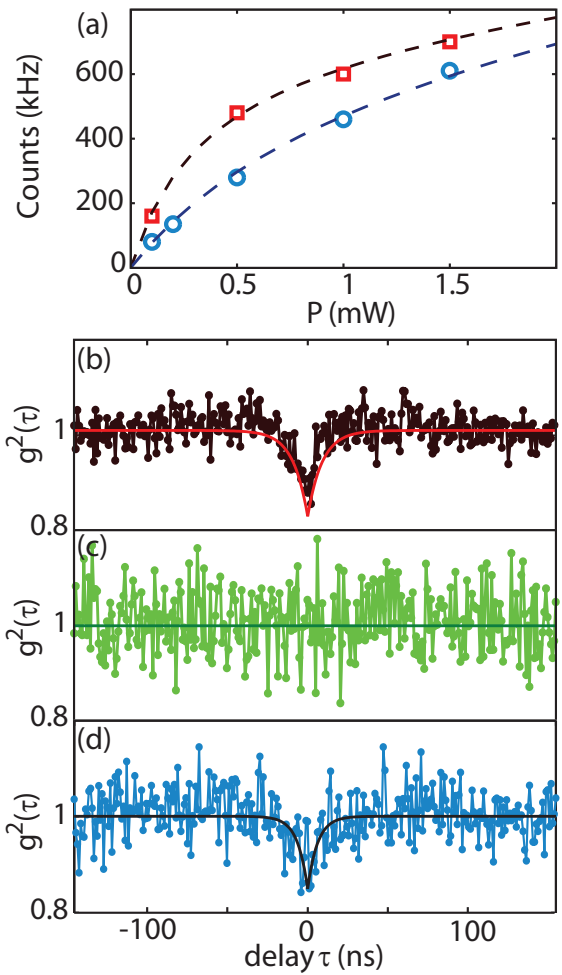}
\end{tabular}
\end{center}
\caption{(a) Saturation curves of the emitted intensity \emph{versus}  excitation power at the tip apex for the NV-based NSOM probe placed in front of the fused silica (red squares) and gold (blue circles) parts of the sample, respectively. These data are obtained in the direct and Fourier spaces, respectively. Dashed curves are theoretical fits (see text).(b-d) Second-order time-intensity correlation function $g^{(2)}(\tau)$ as a function of the time delay $\tau $ between two consecutive detection events in three different detection configurations: (b) fused silica with detection in the direct plane; (c) same as (b) for the gold part of the sample; (d) same as (c) but with detection in the Fourier plane along the SP circle, see Figure 2(d). Red and black curves in (b) and (d) are theoretical fits (see text).}
\end{figure}
\section{Quantum plasmonics and second order correlation measurements using leakage radiation microscopy }
\indent The above analysis shows that we have a solid understanding of the wave-like behavior of leaky SPs launched by the NV-based tip. It is worth emphasizing that leakage radiation is a coherent process in which the cooperative oscillations of the electron cloud on the metal film add coherently to form a well defined SP ring in the Fourier space.~\cite{Drezet2} Therefore, the very good agreement with theory tends to confirm that the first-order spatial coherence of SPs is well preserved during the propagation along the film. This agrees with previous reports, see e.g. ref.~\cite{Bellessa}. In this context we also point out that the broad fluorescence spectrum of the NV centers (see Fig.~5) implies a theoretical first-order spatial coherence of typically $\bar{\lambda}^2/\Delta\lambda\simeq4$ $\mu$m, which is slightly larger that  $L_{SP}$. This in part explains why the agreement between theory and experiment is so good in Fig.~3 and subsequently confirms that the SP ring observed here is essentially a first-order coherent effect. \\
\indent We now turn to the particle-like behavior of SPs. For this purpose, we use an HBT correlator~\cite{Sonnefraud,QM} to measure the second-order time-intensity correlation function $g^{(2)}(\tau)$ \emph{versus} time delay $\tau$ between two emitted SPs. $g^{(2)}(\tau)=P(\tau|0)/P(0)$ measures the conditional probability per unit time $P(\tau|0)$ for recording a single photon (plasmon) at time $\tau $ knowing than one has been previously recorded at time $\tau=0$ (P(0) is the constant single photon detection rate).\cite{QM} Figure 4(b) shows $g^{(2)}(\tau)$ recorded with the NV-based tip facing the fused silica part of the sample. The antibunching dip at zero delay seen in Figure 4(b) is characteristics for 2-level quantum emitters. To analyze quantitatively this antibunching curve, we use the formula \begin{equation}g^{(2)}(\tau)=1-\frac{\rho^2}{N}e^{-\gamma\tau}\end{equation} valid for a ND hosting $N$ identical NV centers.~\cite{Cuche3,NEW} Here, $\gamma= R+1/T$ (R is the pumping rate, $T$ the emitter lifetime) and $\rho=S/(S+B)$ ($S$ is the NV fluorescence signal, $B$ the background intensity coming from the incoherent tip and gold fluorescence). Clearly, it is this lifetime $T$ which could potentially be modified by the metal environment, for example as a result of a change in the local density of states in the vicinity of the interface and an increase of radiative and non-radiative losses related to intrinsic Ohmic processes and to dissipation at the grain boundaries in the rough metal. Here, the best fit (red curve) yields $\frac{\rho^2}{N}=0.18$ and $\gamma=0.11$ ns$^{-1}$. To go further in the analysis requires knowing the saturation behavior of the emitted power $I(P)$, where $I$ is the counting rate and $P$ the excitation power: see Figure 4(a) (red squares). We use the formula~\cite{Novotny} $I=I_{\textrm{sat}}P/(P+P_{\textrm{sat}})+\eta P$ where $P_{\textrm{sat}},I_{\textrm{sat}}$ are the saturation constants of the 2-level system and $\eta$ a constant parameter characterizing the spurious fluorescence. We obtain $I_{\textrm{sat}}=720$ kHz, $P_{\textrm{sat}}=0.34$ mW and $\eta=80$ kHz/mW. Moreover, up to a multiplicative constant, we have $S=I_{\textrm{sat}}P/(P+P_{\textrm{sat}})$ and $B=\eta P$. Hence, at the working power $P=0.1$ mW, where the antibunching curve of Figure 4(b) has been recorded, we have $\rho=0.95$. From this, we obtain the number of NV centers hosted by the elected ND to be $N=5$. Additionally, in this non saturated regime, $R$ can be neglected and we deduce $T_{\textrm{silica}}=9$ ns, in good agreement with the existing literature.~\cite{Gruber} \\
\begin{figure}[h]
\begin{center}
\begin{tabular}{c}
\includegraphics[width=7.5cm]{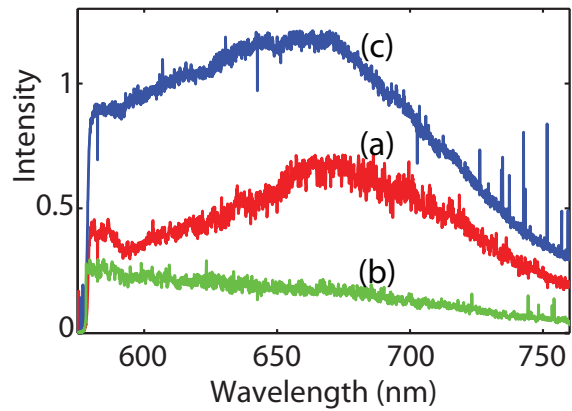}
\end{tabular}
\end{center}
\caption{Fluorescence spectra for the NV-based NSOM tip facing the fused silica substrate (a) or the gold film (b), both acquired in plane 1, and for the NSOM tip facing the gold film, but acquired in plane 2 (c). }
\end{figure}
\indent Figure 4(c) depicts the $g^{(2)}(\tau)$ function recorded with the NV-based tip facing the gold side of the sample. No photon antibunching can be observed whatsoever. However, concluding from this deceptive result that the second-order correlation of the diamond tip would be lost after propagation along the plasmonic film would be erroneous. The real-space signal that is here analyzed includes a large incoherent signal mostly due to gold fluorescence that masks the coherence of the useful signal. This is confirmed by the spectral analysis of the involved signals. Figure 5(a) shows the spectrum recorded in the \emph{direct space} for the NV-based tip placed in front of the fused silica substrate. Clearly, this spectrum shows the usual NV$^{-}$ (i.e. negatively charged NV) signature~\cite{Gruber} which, in turn, is hardly distinguishable when the tip is facing the gold film, as shown in the \emph{direct-space} image in Figure 5(b). In this last configuration, the detected signal is dominated by the gold fluorescence, which is also the reason for the intense central peak seen in the Fourier space LRM images in Figures 2(a) and 2(b). In addition, the detection area in the direct space, shown as a small circle in Figure 2(c), is too small to discriminate between the useful leaky SPs and spurious photons. With the SP signal being delocalized over a large area (see Figure 2(c)), i.e., outside of the fiber detection zone, it is not possible to detect the NV fluorescence in this configuration, which is mainly coupled to leaky SPs. \\
\indent The above discussion appeals to working in the \emph{Fourier space} where spatial filtering can be applied to mitigate the spurious signals. Adding an opaque disk of optimal extension in the back focal plane of the microscope masks almost entirely the allowed light, which has a non-plasmonic origin, whereas the forbidden plasmonic light is unaffected. This is confirmed by the filtered LRM image in Fourier space shown in Figure 2(d) compared to the unfiltered image of Figure 2(b). Now, the detection area for spectral and photon-statistics analysis can be properly positioned on the SP circle to embrace the useful signal only. This is achieved by using a large-core (365 $\mu$m) multimode fiber indicated by the white circle in Figure 2(d). The effect on the measured signal is dramatic as shown in Figure 5(c): the NV spectral signature can be recognized, although it is slightly distorted compared to the original spectrum acquired in front of the silica substrate, possibly as a consequence of the wavelength variation in the coupling of the NVs with the SP modes.~\cite{Marty}\\
\indent Working in the Fourier space places ourselves in a comfortable position to probe the correlation properties of the SP signal generated by the NV-based tip facing the gold film. The HBT correlator is fed with the spatially filtered signal previously analyzed in the spectral domain. The corresponding $g^{(2)}(\tau)$ function, shown in Figure 4(d), now clearly exhibits an antibunching dip. Using the saturation curve shown in Figure 4(a) (blue circles) and repeating the same analysis as before, we obtain $I_{\textrm{sat}}=740$ kHz, $P_{\textrm{sat}}=1$ mW and $\eta=100$ kHz/mW (see fit on Figure 4(a)). From $\rho^2/N$=0.15 and $\rho=0.87$ we deduce $N=5$ just as before, when the tip was facing the silica substrate. This demonstrates that all five NV centers contribute to the SP signal and that the second-order coherence of the initial quantum source of light is not altered by the SP conversion into the gold film, despite the metal roughness. We also obtain $\gamma= 0.16$ ns$^{-1}$ which corresponds to a reduced lifetime $T_{\textrm{metal}}\simeq 6$ ns. This marginal reduction in lifetime, which almost falls within the error bar~\cite{footnote}, could be ascribed to the self-interaction of the quantum emitters close to the metal, i.e. due to a modification of the local density of states.~\cite{GG,Greffet}\\
\indent Before concluding, it is worth stressing how powerful working in the Fourier plane is. It has been shown that Fourier filtering leads to strong image distortions in the direct space.~\cite{Cuche3,Mollet} Therefore, working in the direct space as shown in Figure 2(c), but with the Fourier filter in place, would artificially reduce the depth of the antibunching dip, thereby precluding to conclude about the exact number of contributing NVs. This complication is avoided by working directly in the Fourier space.\\
\section{Summary and conclusions}
\indent To summarize, we have demonstrated for the first time non-classical second-order antibunching $g^{(2)}(0)<1$ through a 2D plasmonic channel. This phenomenon is the quantum counterpart of spontaneous emission in a 2D plasmonic environment. Within this context, LRM plays a major role since it allows for monitoring both the wave-like and particle-like features of quantum SPs emitted by NV centers, thereby highlighting the fundamental wave-particle dualism of quantum plasmons. The high first-order spatial coherence of SPs observed both in the direct and Fourier spaces is a direct signature of their wave-like nature. We plan in the future to study more directly these coherence effects in double-slit-like interference experiments involving SPs and NDs. \\
\indent Finally, the second-order correlation properties of the initial quantum source of light are found to be preserved after SP propagation along a plain film. As a rewarding consequence, it is possible to envision designing lithographically fabricated plasmonic devices for future quantum information processing or cryptography at the nanoscale. The unique potentialities offered by a NSOM system to position very accurately a quantum emitter in a complex plasmonic environment will be in this context a precious advantage to study how this environment modifies both the first- and second-order coherence of the locally emitted photons.     \\
\begin{acknowledgments}
\indent We are grateful to Guillaume Bachelier and James Hulse for helpful discussions and a critical reading of the manuscript, and to Jean-Francois Motte for the optical tip manufacturing. This work was supported by Agence Nationale de la Recherche, France, through the PLASTIPS and NAPHO projects.
\end{acknowledgments}


\begin{thebibliography}{99}
\bibitem{footnote}
Note that the fluctuations in the second-order signal increase from Fig.~4(b) to Fig.~4 (d). This is due to 80$\%$ reduction in the number of detected photons during the full integration time of the correlation histograms when passing from quartz to metal.
\end{thebibliography}
\end{document}